# THE SHER-HIAF RING LATTICE DESIGN

X. Gao*  J.C. Yang  J.W.Xia  W.P. Chai  J. Shi  G.D. Shen
*Institute of Modern Physics, Chinese Academy of Science,
Lanzhou, 730000, Gansu, China
*University of Chinese Academy of Sciences, Beijing, 100049, China

*Abstract*

Super Heavy Experimental Ring (SHER) is one of the rings of the next accelerator complex High Intensity Heavy Ion Accelerator Facility (HIAF) at IMP[4]. Here, present ideas of the lattice design for the operation of the large acceptance ring are presented. The SHER ring has to be optimized for e-cooling and the lattice is designed for different modes. First of all, it is designed in the so called isochronous mode as time-of-flight mass spectrometer for short-lived secondary nuclei. Secondly, SHER can also be used to be a storage ring for collecting and cooling the secondary rare isotope beams from the transport line. In order to fulfil it's purpose, the ion optics can be set to different ion optical modes.

## INTRODUCTION

The present lattice consists of 40 identical 9° sector magnets and 9 quadrupole families (40 quadrupoles in total) to fulfill the first order focusing conditions. The lattice of the SHER consists of two 180 degree arcs separated by two straight lines, it consists of four identical parts, which can decrease the errors caused by fringing field. SHER ring is an achromatic magnetic spectrometer with a momentum acceptance of ±0.45% and a transverse acceptance of 30 πmm·mrad for the transition energy $\gamma_{tr}$=1.835 within the acceptable physical aperture. Heavy-ion beams with magnetic rigidities from 3.2T·m to 25T·m can be analyzed by the facility. The mass-to-charge ratio m/q of the stored ions circulating in the ring can be measured from the revolution time(T) and the velocity (v) of the ions.

$$\frac{\Delta T}{T} = \frac{1}{\gamma_{tr}^2} \cdot \frac{\Delta(m/q)}{(m/q)} + (-1 + \frac{\gamma^2}{\gamma_{tr}^2})\frac{\Delta v}{v} \ .$$

where $\gamma$ is the relativistic Lorentz factor and the $\gamma_{tr}$ is the transition energy[2]. The isochronous condition is reached when $\gamma = \gamma_{tr}$. And the revolution time of the circulating ions is measured with the TOF detector installed in the straight section of the ring. Therefore, measurements of the revolution times give the possibility to determine m/q ratios of circulating ions[2].

## Super Heavy Experimental Ring (SHER)

### Machine Parameters and Lattice

Due to the HIAF design target, the radius of SHER ring should be the same as the former booster ring in order to be upgraded in the future[4]. Different from the present isochronous ring CSRe in IMP, the present layout of the SHER consists of two arcs, where the envelops in the arcs are characterized by a large dispersion function inside the central dipole magnets, in order to achieve the necessary large values of the local momentum compaction factor $\alpha_p$.[3] And if we want to get a small energy transition gamma-t, a large momentum compaction is needed, but a large $\alpha_p$ means the path length varies by a large amount with a small momentum. Many of the aspects are optimized. First of all, the SHER should be operated in the isochronous mode in higher orders. Secondly the transverse and momentum acceptance will be larger than CSRe. Thirdly, the SHER will be achromatic at its straight sections, like CR (in GSI)[1,2], two TOF detectors will be installed in one of the straight sections. In Table 1 the modified parameters of the SHER ring is given. Figure 1 is the layout of the SHER.

Table 1: Modified parameters of the SHER ring

| Mode | ISO | Normal |
|---|---|---|
| Circumference, C, m | 273.3 | 273.3 |
| Number of quadrupoles | 40 | 40 |
| Transition energy, $\gamma_{tr}$ | 1.83 | 3.41 |
| Tune Qx/Qy | 2.36/2.72 | 4.15/2.07 |

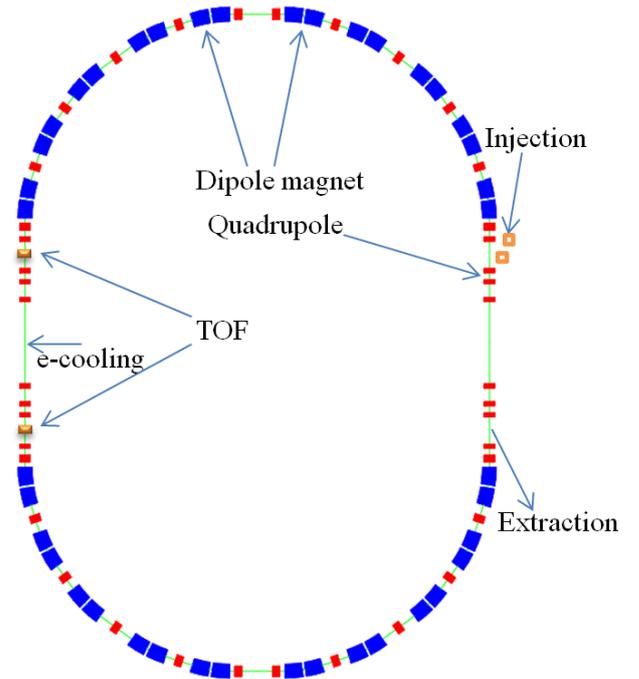

Figure 1: Layout of the SHER

Figure 2 shows the twiss function of the SHER ring with isochronous mode, and figure 3 is the twiss function of normal mode.

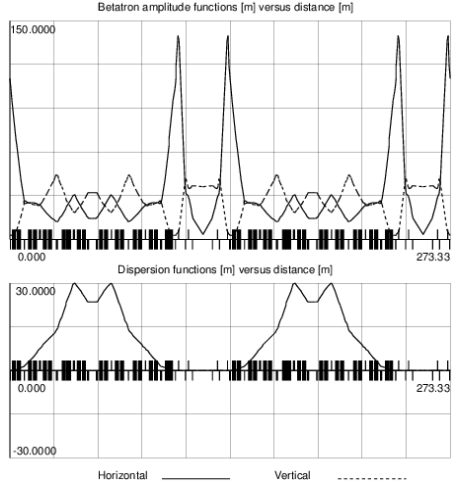

Figure 2: Twiss function of the SHER ring of the first mode with the gamma-t=1.835

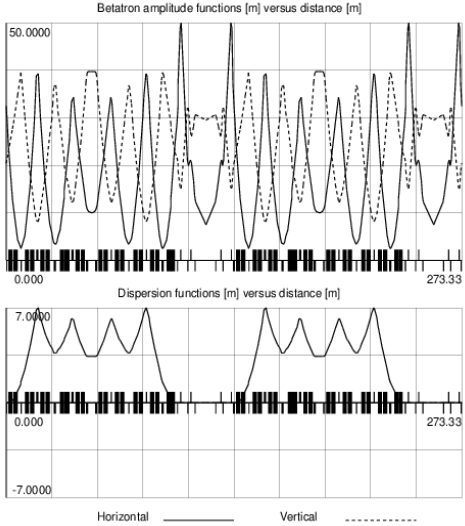

Figure 3: Twiss function of the SHER ring of the normal mode

*Chromatic Correction*

Chromatic correction is also an important part of the SHER ring design. Since the momentum spread of the particles is relatively large (3%) each individual quadrupole give an essential contribution to the chromatic effect of the ring. The focal strength can be written as [3]:

$$f^{-1} = \frac{e(G+G'D\delta)L}{p_0 c(1+\delta)}.$$

where $f^{-1}$ is the focal strength of a quadrupole, G is the quadrupole field component, and $G'$ is the sextupole component, $p_0$ is the momentum of the reference particle, $\delta = \Delta p/p$.

The quadrupole is achromatic if there is a solution of

$$\frac{d}{d\delta}(f^{-1}) = f^{-1}$$

usually, $\frac{\Delta p}{p} < 1$, therefore a quadrupole lens is almost achromatic if

$$G'/G \approx 2/D$$

For SHER, the maximum dispersion in the arc is about 30, the average quadrupole strength is about $0.15 T/m^2$, so the natural chromaticity is small, which is an advantage for dynamitic aperture.

The betatron tunes(Qx,Qy) define the working point of the ring. The tune diagram for SHER is shown in figure 4. Due to the momentum spread the working point is enlarged in the tune diagram and can thus cross a resonance, which can cause the loss of the beam.

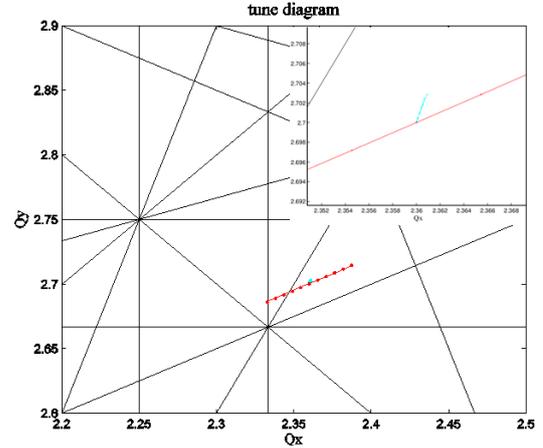

Figure 4: The tune diagram for SHER, the resonance up to 5th order are included. Red and blue curves correspond to the working lines without, with chromaticity correction. The considered momentum spread is $\Delta p/p = \pm 0.5\%$.

*Dynamic Aperture*

The improvement of dynamic aperture starts from optimization of linear optics and chromaticity correction schemes[5]. In the following sections, we review the modifications, and all the tracking studies have been done by Winagile and MAD codes. As an example the tracking results are calculated by the MADX code are shown in Fig.5

Table 2. Summary of major parameters

| Max. mag. rigidity, Bρ,Tm | 20 | |
|---|---|---|
| Circumference, C,m | 273.3 | |
| Super periodicity | 2 | |
| Lattice type | FODO | |
| | RIB | ISO |
| Hor. acceptance, πmm mrad | 70 | 30 |
| Ver. acceptance, πmm mrad | 30 | 30 |
| Mom. acceptance, Δp/p, % | 3 | 0.9 |

The figures show the boundary of stability in the region within which particles survived 1024 turns. and the required horizontal ring acceptance of 30 mm mrad is much smaller than the stable region. Table 2 is the summary of major parameters.

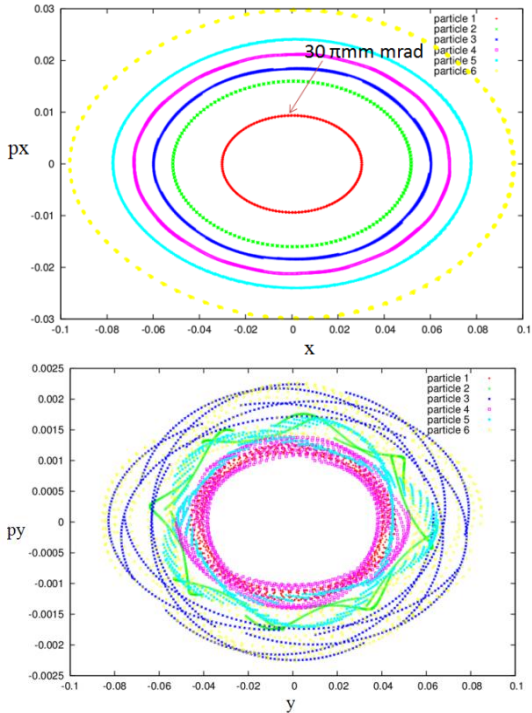

Figure 5: Dynamic aperture of the SHER with isochronous mode optics.

*Injection and Extraction*

The SHER ring is a high acceptance ring with full aperture injection and extraction kickers, e-cooling system. The injection kickers should guarantee that, the full ring acceptance is available for the incoming hot secondary beams[1,5]. For SHER, the injection and extraction should be in the same straight line, so kickers are installed in the middle of the straight section, for both injection and extraction. The trajectories of the injection beams between the first septum and the last injection kicker magnet are shown in Fig.6. The beam envelope in the septum is 82.1mm(for the collect mode), and 74.3mm(for the isochronous mode), so the septum should be placed at 126.8mm, with COD, clearance, thickness considered. The injection septum is placed between two quadrupoles, where the phase advance between the injection septum and the injection kicker amounts to about 90°. Six injection kicker modules are needed to produce a kicker angle of 12mrad. And the parameters for the septum magnet are shown in table 3.

Table 3. Parameters for the septum magnet

| Length, m | 1.8 |
|---|---|
| Horizontal angle, degree | 9 |
| Dipole bending field, T | 1.309 |

| Physical aperture, mm・mm | 110×60 |
|---|---|

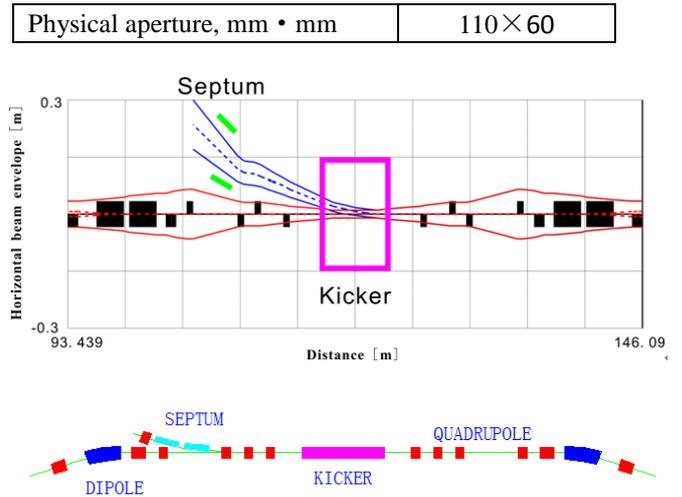

Figure 6: Injection schemes of the beam for SHER